\documentstyle[prl,aps]{revtex}
\draft

\begin{document}    

\title{
\vskip-3cm{\baselineskip14pt
\centerline{\normalsize\hfill BUTP--99/15}
\centerline{\normalsize\hfill TTP99--33}
\centerline{\normalsize\hfill hep-ph/9907509}
\centerline{\normalsize\hfill July 1999}
}
\vskip.7cm
Short Distance Mass of a Heavy Quark at Order $\alpha_s^3$
\vskip.3cm
}
\author{
{K.G. Chetyrkin}\thanks{Permanent address:
Institute for Nuclear Research, Russian Academy of Sciences,
60th October Anniversary Prospect 7a, Moscow 117312, Russia.}}
\address{Institut f\"ur Theoretische Teilchenphysik,
  Universit\"at Karlsruhe, D-76128 Karlsruhe, Germany}
\author{M. Steinhauser}
\address{
  Institut f\"ur Theoretische Physik,
  Universit\"at Bern, CH-3012 Bern, Switzerland}
\maketitle

\begin{abstract}
\noindent
The relation between the on-shell quark mass and the mass defined in
the modified minimal subtraction scheme is computed up to order
$\alpha_s^3$. Implications for the numerical values  of the top
and bottom quark masses are discussed.  We show that the new
three-loop correction  significantly reduces the
theoretical uncertainty in the determination of the 
quark masses.

\end{abstract}


In Quantum Chromodynamics (QCD)
practical calculations are very often performed in the
modified minimal subtraction ($\overline{\rm MS}$)
scheme~\cite{tHo73,BarBurDukMut78}
leading to the definition of the so-called short-distance $\overline{\rm MS}$
mass.
The $\overline{\rm MS}$ mass occupies a distinguished place among
various mass definitions. First, it is a truly short distance mass not
suffering from nonperturbative ambiguities. Second, the $\overline{\rm
MS}$ mass proves to be extremely convenient in multi-loop calculations
of mass-dependent inclusive physical observables dominated by short
distances (for a review see~\cite{ckk96}). 
On the other hand the experiments often provide masses
which are tightly connected to the on-shell definition.
Thus, conversion formulae
are needed in order to make contact between theory and experiment.
The two-loop relation between $\overline{\rm MS}$ and the on-shell definition
of the quark mass has been obtained in~\cite{GraBroGraSch90}.
Until recently the accuracy of this equation was enough for the practical
applications.
Meanwhile, however, new computations have become available which require the
relation between the $\overline{\rm MS}$ and on-shell mass at ${\cal
  O}(\alpha_s^3)$ in order to perform a consistent analysis. 
The necessity of an accurate determination of the quark masses,
especially those of the top and bottom ones,
is demonstrated by the following two examples:

$(i)$ The main goal of the future $B$ physics experiments is the
determination of the Cabibbo-Kobayashi-Maskawa matrix elements which will
give deeper insight into the origin of CP violation and possibly also 
provides hints to new physics.
In particular the precise measurement of $V_{cb}$ is very promising.
It is determined from semileptonic $B$ meson
decay rates. Thus it is desirable to know the bottom quark mass
as accurately as possible as it enters already the Born result to the fifth
power.

$(ii)$ One of the primary goals of a future electron-positron linear collider
(NLC) or muon collider (FMC) will be the precise determination of the top
quark properties, especially its mass, $M_t$.
In hadron colliders like the Fermilab TEVATRON or
the Large Hadron Collider (LHC) the top quarks are reconstructed from
the invariant mass of the $W$ bosons and the bottom quarks.
On the contrary in lepton colliders
it is possible to determine the top quark mass from the line shape of the
production cross section $\sigma(e^+ e^- \to t\bar{t})$ close to the
threshold.
Simulation studies have shown that an experimental uncertainty
of $100-200$~MeV in the top mass determination can be
achieved~\cite{sim95}.
Thus also from the theoretical side the ambiguities have to be
controlled with the same precision.
In particular in this context much attention has been devoted to the relation
of the pole mass, $M$, to the $\overline{\rm MS}$ mass, $m$. Although the pole
mass demonstrates a bad infra-red behaviour it is often convenient to use it
in intermediate steps.

The connection between the $\overline{\rm MS}$ and on-shell mass
is given by
\begin{eqnarray}
m(\mu) &=& z_m(\mu) M
\,,
\label{eq:mmsos}
\end{eqnarray}
where $z_m$ is finite and has an explicit dependence on the
renormalization scale $\mu$.
In~\cite{Kro98} the infra-red finiteness and
the gauge invariance of $M$ was proven. 
In~\cite{GraBroGraSch90} 
the perturbative expansion of $z_m$ has been computed up to order
$\alpha_s^2$. 
The main purpose of this letter
is the computation of $z_m$ up to order $\alpha_s^3$.
Therefore three-loop corrections to the fermion propagator have to be
considered.
It is convenient to parameterize them in the following form
\begin{eqnarray}
  z_m(M) &=& 1 + a_M C_F z_m^F
  + a_M^2
  \Bigg[
    C_F^2 z_m^{FF} + C_FC_A z_m^{FA} 
    + C_FTn_l z_m^{FL} + C_FT z_m^{FH}
  \Bigg]
  + a_M^3\,z_m^{(3)}
  +{\cal O}\left(a_M^4\right)
  \,,
  \label{eq:deccf}
\end{eqnarray}
with $a_M=\alpha_s^{(n_f)}(M)/\pi$.
$n_l$ is the number of light (massless) quarks.
In the following we will use $z_m=z_m(M)$.

The relation between the $\overline{\rm MS}$ and on-shell mass is obtained
from the requirement that the inverse fermion propagator has
a zero at the position of the on-shell mass.
Thus in principle it is necessary to evaluate three-loop
on-shell integrals in order to obtain the ${\cal O}(\alpha_s^3)$ term in the
$\overline{\rm MS}$--on-shell mass relation.
This is avoided by computing
expansions of the quark self energy for small and large external
momentum.
After building the proper combinations needed for the relation between the
$\overline{\rm MS}$ and on-shell mass
a conformal mapping $q^2/M^2=4\omega/(1+\omega)^2$ is performed
which maps the complex $q^2$-plane into the interior of the unit circle in
the $\omega$-plane. The relevant point $\omega = 1$ (corresponding
to $q^2=M^2$) is reached with the help of Pad\'e
approximation~\cite{FleTar94}.
For further details we refer the reader to~\cite{CheSte99_long}.
Note that this method has already been extremely successful in other
applications~\cite{pade}.

Let us at this point discuss the one- and two-loop results
in order to get some feeling about the quality of our procedure.
Using our method the following numbers can be extracted
\begin{eqnarray}
\begin{array}{lllll}
\displaystyle
z_m^{F} \,\,=\,\, -1.0005(6)
\,,
&
\displaystyle
z_m^{FF} \,\,=\,\, -0.49(2)
\,,
&
\displaystyle
z_m^{FA} \,\,=\,\, -3.4(1)
\,,
&
\displaystyle
z_m^{FL} \,\,=\,\, 1.566(4)
\,,
&
z_m^{FH} \,\,=\,\, -0.1553(2)
\,,
\label{eq:zm12}
\end{array}
\end{eqnarray}
where the error is obtained by doubling the  spread of the
different  Pad\'e approximants.
The comparison with the exact result results~\cite{GraBroGraSch90}
$\{-1$, $-0.51056$, $-3.33026$, $1.56205$, $-0.15535\}$ shows 
very good agreement.

The results obtained at one- and two-loop order encourage to apply the same
procedure also at order $\alpha_s^3$.
For the technical details we also refer to~\cite{CheSte99_long}.
We just want to mention that in total we were able to evaluate 14 input
terms for the Pad\'e procedure.

We could apply the described procedure to each colour factor
separately~\cite{CheSte99_long}.
For most practical applications in QCD the number of flavours may be set to 3,
i.e. $C_F=4/3$ and $C_A=3$, and $T=1/2$ can be adopted.
Adding in a first step the results for the moments and performing
the Pad\'e approximations afterwards leads us to
\begin{eqnarray}
  \frac{m(M)}{M} &=&
  1 
  -1.333 
  a_M
  +a_M^2
  \left[
    -14.33 + 1.041 n_l
  \right]
  +a_M^3
  \left[
    -202(5) + 27.3(7) n_l  - 0.653 n_l^2
  \right]
  \,,
  \label{eq:zmlog2}
  \\
  \frac{\mu_m}{M} &=&
  1 
  -1.333
  a_M
  +a_M^2
  \left[
    -11.67 + 1.041 n_l
  \right]
  +a_M^3
  \left[
    -170(5) + 24.8(7) n_l - 0.653 n_l^2
  \right]
  \,,
  \label{eq:zmzm2}
  \\
  \frac{M}{m(m)} &=&
  1 
  +1.333
  a_m
  +a_m^2
  \left[
    13.44 - 1.041 n_l
  \right]
  +a_m^3
  \left[
    194(5) - 27.0(7) n_l + 0.653 n_l^2
  \right]
  \,,
  \label{eq:zminv2}
\end{eqnarray}
with $a_m=\alpha_s^{(n_f)}(m)/\pi$.
For simplicity $\mu=M$ and $\mu=m$ has been chosen
in~(\ref{eq:zmlog2}) and~(\ref{eq:zminv2}), respectively.
In the above equations the exact values of the $n_l^2$ term~\cite{BenBra95}
is displayed. Our method leads to 0.656(8) which is in very good agreement.
This is a further justification for our approach.

The results for the values of $n_l=0,\ldots,5$ are summarized in
Tab.~\ref{tab:nl} where also the two-loop coefficients are displayed.
Actually the coefficients of the terms linear in $n_l$ have been
obtained by performing a fit to the three-loop results of
Tab.~\ref{tab:nl}.  The  errors  of about  2--3\%  
for the three-loop results 
of~(\ref{eq:zmlog2})--(\ref{eq:zminv2}) and Tab.~\ref{tab:nl}
have again been obtained by doubling the spread of Pad\'e approximants.
On one side this is justified with the
behaviour at ${\cal O}(\alpha_s^2)$ where the results of the first
column in Tab.~\ref{tab:nl} are reproduced with an accuracy below 2\%.
On the other side the Pad\'e approximants demonstrate more
stability in the case where the moments are added and
$n_l$ is fixed afterwards
than in the case where the Pad\'e procedure is applied to the
individual colour structures separately.
We want to stress that the errors in Eqs.~(\ref{eq:zmlog2})--(\ref{eq:zminv2}) 
are somewhat overestimated once a value for $n_l$ is chosen.
Thus for practical applications Tab.~\ref{tab:nl} should be used.

Let us next compare our results with various
predictions which already exist in the literature
obtained with the help of different optimization procedures.
In Tab.~\ref{tab:FACPMS} the results obtained for $M/m(m)$
using the fastest apparent convergence (FAC)~\cite{FAC}
and the principle of minimal sensitivity (PMS)~\cite{PMS}
are compared with ours.
For $n_l=2$ the discrepancy with our central value
amounts to only 7\%. It even reduces to
2\% for $n_l=5$, i.e. in the case of the top quark.
The predictions obtained in the large-$\beta_0$ limit,
where $\beta_0$ is the first coefficient of the QCD $\beta$ function,
are also shown in Tab.~\ref{tab:FACPMS}.
Excellent agreement below 1\% is found for $n_l=3$. It amounts to
roughly 5\% for $n_l=4$ and 14\% for $n_l=5$.

In the remaining part of this letter we will discuss some important
applications of the new term of ${\cal O}(\alpha_s^3)$
in the $\overline{\rm MS}$--on-shell relation.

Threshold phenomena are conveniently expressed in terms of 
the pole mass.
To be specific let us consider the production of top quarks in
$e^+e^-$ collisions.
The corresponding physical observables
expressed in terms of $M_t$ show in general a bad
convergence behaviour. In the case of the total cross section the
next-to-next-to-leading order corrections~\cite{ttNNLO,BenSigSmi99,HoaTeu99}
partly exceed the next-to-leading
ones. Furthermore the peak position which is the most striking feature of the
total cross section and from which finally the mass value
can be extracted depends strongly
on the number of terms one includes into the analysis.
The reason for this is that
the pole mass is sensitive to long-distance effects which results in
intrinsic uncertainties of order $\Lambda_{\rm QCD}$~\cite{BenBra94}.

Several strategies have been proposed to circumvent this
problem~\cite{Ben98,HoaSmiSteWil98,HoaTeu99}.
They are based on the observation that
the same kind of ambiguities also appear in the static quark
potential, $V(r)$. In the combination $2M_t + V(r)$, however, the
infra-red sensitivity drops out. Thus a definition of a short-distance
mass extracted from threshold quantities should be possible.
The relation of the new mass parameter
to the pole mass is used in order to re-parameterize the threshold phenomena.
On the other hand a relation of the new quark mass to the $\overline{\rm MS}$
mass must be established as
it is commonly used for the parameterization of those quantities which are not
related to the threshold.
In order to do this consistently
the three-loop relation between the
$\overline{\rm MS}$ and the on-shell mass is needed.

In~\cite{Ben98} the concept of the so-called potential mass, $m_{t,PS}$, has
been introduced. Its connection to the pole mass is given by
$m_{t,PS}(\mu_f) = M_t - \delta m_t(\mu_f)$
where $\delta m_t(\mu_f)$ is obtained from the static quark potential.
In this way a subtracted potential, $V(r,\mu_f)$, is defined.
The factorization scale $\mu_f$ has been introduced in order to extract
the infra-red behaviour arising from the potential.
In the combination $M_t - \delta m_t(\mu_f)$
it cancels against the one of $M_t$ leading to an significant 
reduction of the long-distance uncertainties in $m_{t,PS}$~\cite{Ben98}.
Thus it is promising to formulate the threshold problems in terms of
$m_{t,PS}(\mu_f)$ and $V(r,\mu_f)$ instead of $M_t$ and $V(r)$.
In the final result the dependence on $\mu_f$ cancels.
For the numerical analyses the value $\mu_f=20$~GeV has been adopted
in~\cite{BenSigSmi99} as its upper bound is roughly given by
$M_tC_F\alpha_s(\mu)$.
$\delta m_t(\mu_f)$ is known up to order $\alpha_s^3$~\cite{Ben98}.
Thus we are now in the position to establish a relation between
the two short-distance masses $m_{t,PS}(\mu_f)$ and $m_t(\mu)$
with the result:
\begin{eqnarray}
  m_{t,PS}(20~{\rm GeV}) &=& \left( 165.0 + 6.7 + 1.2 + 0.28
  \right)~\mbox{GeV}
  \,,
  \label{eq:mtPSmtmt}
\end{eqnarray}
where the different terms represent the contributions of order $\alpha_s^0$
to $\alpha_s^3$. For the numerical values
$m_t(m_t)=165.0$~GeV and $\alpha_s^{(6)}(m_t(m_t))=0.1085$
have been used.
Note that the error of the ${\cal O}(\alpha_s^3)$ coefficient
in the $\overline{\rm MS}$--on-shell mass relation is negligible.
The comparison of Eq.~(\ref{eq:mtPSmtmt}) with the analogous expansion
for $M_t$,
\begin{eqnarray}
  M_t &=& ( 165.0 + 7.6 + 1.6 + 0.51 )\mbox{~GeV}
  \,,
\end{eqnarray}
shows that the potential mass can be 
more accurately related to the $\overline{\rm MS}$ mass than $M_t$.
The last term of the expansion in~(\ref{eq:mtPSmtmt})
is of the same order of magnitude as the error in the top quark mass
determination at a NLC.
Whereas in~\cite{BenSigSmi99} this term 
has been taken as uncertainty in the mass relation
the error reduces significantly after the knowledge of the
${\cal O}(\alpha_s^3)$ term of the $\overline{\rm MS}$--on-shell relation.
This can be deduced from the well-behaved expansion in
Eq.~(\ref{eq:mtPSmtmt}).
The dominant error is now provided by the uncertainty in $\alpha_s$.

A similar strategy has been proposed in~\cite{HoaTeu99}.
There the so-called 1S~mass, $M_t^{1S}$, has been
defined as half the perturbative
mass of a fictious toponium $1^3S_1$ ground state
which would exist if the top quark were stable.
The philosophy is very similar as in the case of the potential mass.
From the experiment the quantity $M_t^{1S}$ is extracted.
In~\cite{HoaTeu99} it has been shown that this is possible with an uncertainty
of approximately $200$~MeV. In a next step $M_t^{1S}$ has to be related to
the $\overline{\rm MS}$ mass $m_t(m_t)$.
As the extraction of $M_t^{1S}$ is based on a
next-to-next-to-leading order formalism the ${\cal O}(\alpha_s^3)$ relation
computed in this work is necessary.

In the practical calculation care has to be taken in connection to the
expansion parameter which has to be used.
Actually the so-called $\Upsilon$-expansion has to be
adopted. Details can be found in~\cite{HoaLigMan99,Hoa99}.
One finally arrives at the following relation between the
$\overline{\rm MS}$ and $1S$~mass
\begin{eqnarray}
  m_t(m_t) &=& \left( 175.00 - 7.60 - 0.97 - 0.14 \right)~{\rm GeV}
  \label{eq:mtmtM1S}
  \,,
\end{eqnarray}
where $M_t^{1S}=175$~GeV and $\alpha_s^{(5)}(M_Z)=0.118$ has been adopted.
Using the large-$\beta_0$ results for the order $\alpha_s^3$
term the last term reads $-0.23$~\cite{HoaTeu99}
which is off by more than 50\% form the exact result.
The conclusions which can be drawn from Eq.~(\ref{eq:mtmtM1S}) are very
similar to the ones stated above: the uncertainties due to unknown terms in the
mass relations are negligible as compared to the error with which $M_t^{1S}$
can be extracted from the experiment. The dominant uncertainty
comes from the error in $\alpha_s$ which amounts for $\pm0.003$ to roughly
200~MeV~\cite{HoaTeu99} in Eq.~(\ref{eq:mtmtM1S}).

Also the bottom quark mass can be extracted from
quantities related to the quark threshold.
Recently~\cite{MelYel99PenPiv99BenSig99,Hoa99}
a precise value for the bottom quark mass has
been determined in the context of QCD sum rules.
For example, in~\cite{Hoa99} the on-shell mass was eliminated in
favour of the 1S mass in order to reduce the error.  Once $M_b^{1S}$ is
determined, $m_b(m_b)$ can be found in analogy with the top quark case.
As a result the values $M_b^{1S}=4.71\pm0.03$~GeV and
$m_b(m_b)=4.2\pm0.06$ have been obtained.
Following the procedure described in~\cite{Hoa99} one arrives at
\begin{eqnarray}
  m_b(m_b) &=& \left( 4.71 - 0.40 - 0.11 - 0.03 \pm 0.03 \pm 0.04
               \right)~\mbox{GeV}       
  \label{eq:mbmbM1S}
  \,,
\end{eqnarray}
where the different terms correspond to different orders in the
$\Upsilon$-expansion.
The first error is due to $M_b^{1S}$ and the second one reflects
the error in $\alpha_s^{(5)}(M_Z)=0.118\pm0.004$ which is adopted
from~\cite{Hoa99}. Due to the nice convergent behaviour
in~(\ref{eq:mbmbM1S}) the total error on $m_b(m_b)$ only contains these
two sources which finally leads to
\begin{eqnarray}
  m_b(m_b) = 4.17 \pm 0.05~\mbox{GeV}
  \,.
\label{mb(mb)_from_M1S:final}
\end{eqnarray}
It is important to stress that taking into account of the newly
computed ${\cal O}(\alpha_s^3)$ term in the $\overline{\rm
MS}$--on-shell relation is crucial for the reliable estimation of
the errors in (\ref{mb(mb)_from_M1S:final}). Indeed,  a deviation of 
the real value for $z_m^{(3)}$ from the large-$\beta_0$ estimation 
by, say, a factor of two, which one could not exclude a priori,
would result to a systematic shift in  $m_b(m_b)$
of around  100~MeV.

A somewhat different approach for the determination of both the charm and
bottom mass has been followed in~\cite{PinYnd98}.
There the  lower states in the heavy quarkonium spectrum were  computed
up to order $\alpha_s^4$
and then used to extract the pole masses of $b$ and $c$ quarks. 
The transformation to the $\overline{\rm MS}$ mass has been performed with the
help of the two-loop relation~\cite{GraBroGraSch90}.
However, to the order the quarkonium spectrum was computed
it is more consistent to use the ${\cal O}(\alpha_s^3)$ relation provided in
this paper.
Thus taking over the error estimates from~\cite{PinYnd98}
the on-shell value for the bottom quark mass reads
$M_b = 5.001^{+0.104}_{-0.066}\mbox{~GeV}$.
It transforms to the following $\overline{\rm MS}$ value
\begin{eqnarray}
  m_b(m_b) &=& 4.322^{+0.043}_{-0.028}\mbox{~GeV}
  \,,
\end{eqnarray}
where the value $\alpha_s^{(5)}(M_Z)=0.114$ has been used
as in~\cite{PinYnd98}.
Compared to~\cite{PinYnd98} the inclusion of the ${\cal
O}(\alpha_s^3)$ terms leads to a shift in the central values of more
than $100$~MeV.  In the case of the bottom quark this change is even
larger than the errors presented in~\cite{PinYnd98}. This demonstrates
that a consistent treatment of the different orders in $\alpha_s$ is
absolutely crucial. 

To summarize:
the computed value of the next-to-next-to-leading  
correction to the $\overline{\rm MS}$--on-shell mass relation 
proves to be in a good agreement with an estimation based 
on the large-$\beta_0$ limit and  has led to a significant reduction of
the theoretical uncertainty in the determination of the quark masses.


\vspace{.5em}

We would like to thank A.H.~Hoang for useful conversation
and I.~Bigi, J.H.~K\"uhn and K.~Melnikov for
valuable suggestions and careful reading of the manuscript.
This work was supported by DFG under Contract Ku 502/8-1
({\it DFG-Forschergruppe ``Quantenfeldtheorie, Computeralgebra und
  Monte-Carlo-Simulationen''})
and the {\it Schweizer Nationalfonds}.


\begin{table}[b]
  \begin{tabular}{l||r|r|r|r|r|r} 
        & \multicolumn{2}{c|}{$m(M)/M$}
        & \multicolumn{2}{c|}{$\mu_m/M$}
        & \multicolumn{2}{c}{$M/m(m)$}
        \\
        \hline
        $n_l$ 
        & ${\cal O}(\alpha_s^2)$ & ${\cal O}(\alpha_s^3)$
        & ${\cal O}(\alpha_s^2)$ & ${\cal O}(\alpha_s^3)$
        & ${\cal O}(\alpha_s^2)$ & ${\cal O}(\alpha_s^3)$
        \\
        \hline
$0$ &
$    -14.33$ & $   -202(5)$ &
$    -11.67$ & $   -170(5)$ &
$     13.44$ & $    194(5)$ \\
$1$ &
$    -13.29$ & $   -176(4)$ &
$    -10.62$ & $   -146(4)$ &
$     12.40$ & $    168(4)$ \\
$2$ &
$    -12.25$ & $   -150(3)$ &
$     -9.58$ & $   -123(3)$ &
$     11.36$ & $    143(3)$ \\
$3$ &
$    -11.21$ & $   -126(3)$ &
$     -8.54$ & $   -101(3)$ &
$     10.32$ & $    119(3)$ \\
$4$ &
$    -10.17$ & $   -103(2)$ &
$     -7.50$ & $    -81(2)$ &
$      9.28$ & $     96(2)$ \\
$5$ &
$     -9.13$ & $    -82(2)$ &
$     -6.46$ & $    -62(2)$ &
$      8.24$ & $     75(2)$ \\
      \end{tabular}
      \caption{\label{tab:nl}
        Dependence of $z_m^{(2)}$ and $z_m^{(3)}$ on $n_l$. The choice
        $\mu^2=M^2$, respectively, $\mu^2=m^2$
        has been adopted. $z_m^{(2)}$ is defined as the sum of the terms
        inside the square brackets in Eq.~(\ref{eq:deccf}).
        }
  \end{table}

  \begin{table}[b]
      \begin{tabular}{l||r||r|r|r}
        $n_l$ 
        & this work
        & \cite{CheKniSir97} (FAC)
        & \cite{CheKniSir97} (PMS)
        & \cite{BenBra95} (large-$\beta_0$)
        \\
        \hline
        $2$ &
        $    143(3)$
        & $152.71$ & $153.76$ & $137.23$\\
        $3$ &
        $    119(3)$
        & $124.10$ & $124.89$ & $118.95$\\
        $4$ &
        $     96(2)$
        & $97.729$ & $98.259$ & $101.98$\\
        $5$ &
        $     75(2)$
        & $73.616$ & $73.903$ & $86.318$\\
      \end{tabular}
      \caption{\label{tab:FACPMS}Comparison
        of the results obtained in this
        paper with estimates based of FAC, PMS and the large-$\beta_0$
        approximation for $M/m(m)$.
        }
  \end{table}

\end{document}